\documentstyle[11pt]{article}

\begin{document}
\begin{center}
 
{\Large\bf On time-dependent quasi-exactly solvable problems}
\footnote{This work was partially supported by DFG grant No. 436 POL 
113/77/0 (S)}
\vskip 1cm

{\large {\bf Dieter Mayer} }

\vskip 0.1 cm
 
Institute of Theoretical Physics, TU Clausthal,\\
Arnold Sommerfeld Str. 6, 38-678 Clausthal, Germany\footnote{E-mail
address: dieter.mayer@tu-clausthal.de}\\
 
\vskip 1cm

{\large {\bf Alexander Ushveridze}  and {\bf Zbigniew Walczak}}
 
\vskip 0.1 cm
 
Department of Theoretical Physics, University of {\L}\'od\'z,\\
ul. Pomorska 149/153, 90-236 {\L}\'od\'z, Poland\footnote{E-mail
addresses: alexush@mvii.uni.lodz.pl and walczak@mvii.uni.lodz.pl} \\

\end{center}
\vspace{1 cm}
\begin{abstract}
In this paper we demonstrate that there exists a close relationship between
quasi-exactly solvable quantum models and two special classes of classical
dynamical systems. One of these systems can be considered a natural
generalization of the multi-particle Calogero-Moser model
and the second one is a classical matrix model.
\end{abstract}
\thispagestyle{empty}
\newpage
 
\section{Introduction}
 
A quantum mechanical model is called {\em quasi-exactly solvable} (QES) if a
finite number of energy levels and the corresponding wavefunctions of the 
model can be constructed explicitly. One possible way of studying the QES
models and their solutions in the one-dimensional case is based on a
reformulation of the spectral equations in terms of the wavefunction zeros. 
It can be shown that, in all the cases when the number of wavefunction zeros
is finite (and this is just the case of QES models), the problem of 
reconstructing solutions of the stationary Schroedinger equation
becomes purely algebraic and is reduced to the determination of their 
positions in the complex plane. These positions (which hereafter we 
shall denote by $\xi _i)$ obey the system of algebraic equations
\begin{equation}
\sum_{k=1,k\neq i}^M\frac \hbar {\xi _i-\xi _k}+F(\xi _i)=0,\quad 
i=1,\dots ,M
\label{int1}
\end{equation}
where $F(\xi )$ is a rational function in which all the information of a QES
model is contained. In turns out that equations of the type (\ref{int1}) are
typical not only for QES models but appear in many branches of  mathematical
physics and this fact enables one to establish a deep relationship  between
QES models and many other seemingly unrelated models. 
For example, QES  models turn out to be equivalent to completely 
integrable Gaudin spin chains \cite{UshRev,Ush} (for which system 
(\ref{int1}) plays the role of the Bethe ansatz equations for the 
coordinates of elementary spin excitations), 
to random matrix models \cite{Cic1,Cic2} (for which system (\ref{int1})
determines the distribution of eigenvalues of large random matrices), 
and also to purely classical models of 2-dimensional elctrostatics 
\cite{UshRev,UshLeb,Shi} respectively hydrostatics of point vortices 
\cite{May}. For both the latter models system (\ref{int1}) determines 
the equilibrium positions of pointwise classical objects (the charged 
Coulomb particles or resp. vortices) in an external (electrostatic 
or resp. hydrostatic) field.\\
 
Up to now only the stationary solutions of QES Schroedinger equations 
have been considered from this point of view. The aim of the present 
paper is to consider time dependent solutions for QES models and to 
show how they can equivalently be described as dynamical equations for
the motion of wavefunction zeros \footnote{We mean here just the
solutions of evolution equations for standard QES models with
time-independent potentials. Note that QES models with time-dependent
potentials were discussed (from different point of view) in paper
\cite{FinKam}}. 
This enables one to reveal three classical (complex) dynamical
systems closely related to quantum QES models. One of these classical
systems does not have an immediate physical interpretation, but the two 
remaining ones are very interesting from both the physical and 
mathematical point of view. 
The point is that one of these two systems is a natural 
generalization of the famous classical Calogero-Moser multi-particle 
system and the second one is a classical matrix model from which the 
first system can be obtained by means of Olshanetsky and Perelomov's 
projection method \cite{Olsh1}.\\
 
Stricly speaking, the idea of studying classical dynamical systems
describing the motion of zeros of solutions of linear differential 
equations belongs to Calogero and is far from being new. 
Papers devoted to the investigation of such systems appear in the 
literature rather frequently.
The goal of our paper is to apply Calogero and Olshanetsky-Perelomov
methods to 
a concrete class of quantum QES models and to derive the associated classical
models. The most interesting and quite unexpected result which we intend to
present here is that the potentials of the resulting classical matrix 
models turn out to almost coincide with the potentials of the initial 
quantum QES ones (up to terms of order $\hbar $). This fact may hint to the 
existence of a certain non-standard ``quantization procedure'' relating the 
classical multi-particle systems of Calogero-Moser type to QES models of
one-dimensional quantum mechanics.\\
 
The paper is organized as follows: in Section 2 we remind the reader of 
the basic facts concerning the simplest QES model in the stationary case. In
Section 3 we derive the explicit form of solutions for this model in the
time-dependent case and the corresponding system of evolution
equations for the wavefunction zeros. 
In Section 4 we show thatsolutions of this system can be
considered as ``soliton like solutions''  of a multi-particle
classical system of the Calogero-Moser type. The matrix version of 
this system is discussed in Section 5. In Section 6 we discuss the 
limiting case when the number of algebraically calculable states in 
the QES model tends to infinity. The last section is devoted to a 
discussion of our results.
 
\section{The simplest QES model}
 
The simplest QES model is the sextic anharmonic oscillator. Its potential 
has the following form
\begin{equation}
\label{sextic1}
  V(x)=\frac{x^2}2\left(ax^2+b\right)^2-
  \hbar a\left(M+\frac{3}{2}\right) x^2,
\end{equation}
where $a$ and $b$ are real parameters with $a>0$, and $M$ is an arbitrary
non-negative integer. It can be shown \cite{KrajUshWal} that for any 
given $M\geq 0$, the stationary Schroedinger equation
\begin{equation}
\label{sextic2}
  \left(-\frac{\hbar^2}2\frac{d^2}{dx^2}+V(x)\right)\Psi(x)=E\Psi(x)
\end{equation}
for model (\ref{sextic1}) admits solutions of the form
\begin{equation}
\label{sextic3}
  \Psi(x)=\prod_{i=1}^{M}(x-\xi_i)\exp \left[-\frac{1}{\hbar}
  \left(\frac{bx^2}{2}+\frac{ax^4}{4}\right) \right] ,
\end{equation}
\begin{equation}
\label{sextic4}
  E=\hbar b\left(M+\frac{1}{2}\right)+\hbar a\sum_{i=1}^{M}\xi_i^2,
\end{equation}
where the numbers $\xi_i$, $i=1,\dots,M$ (playing the role of the  
wavefuntion zeros) satisfy the system of numerical equations
\begin{equation}
\label{sextic5}
  \sum_{k=1,k\neq i}^{M}\frac{\hbar}{\xi_i-\xi_k}=
  b\xi_i+a\xi_i^3,\quad i=1,\dots,M
\end{equation}
with the following additional condition
\begin{equation}
\label{sextic6}
  \sum_{i=1}^{M}\xi_i=0.
\end{equation}
Note that the form of system (\ref{sextic5}) coincides exactly  with that 
of the famous Bethe Ansatz equations appearing in the theory of completely
integrable Gaudin models \cite{Gaud,Ush}. Therefore we hereafter will 
call (\ref{sextic5}) the Bethe Ansatz equations.
 
It is not difficult to show that for any fixed $M$ the Bethe Ansatz
equations (\ref{sextic5})-(\ref{sextic6}) have
\begin{equation}
\label{sextic7}
  N_{sol}=\left[\frac{M}{2}\right]+1
\end{equation}
solutions. This means that for the sextic anharmonic oscillator we can 
find $N_{sol}$ wavefunctions $\Psi _i(x)$ and $N_{sol}$ corresponding energy
levels $E_i$ by means of purely algebraic methods.\\
 
Note also that model (\ref{sextic1}) can be considered a deformation 
of the simple harmonic oscillator with the potential
\begin{equation}
\label{sextic8}
  V_0(x)=\frac{b^2x^2}{2}.
\end{equation}
The role of the deformation parameter is played by $a$. It is not
difficult to see that after taking $a=0$ in formulas 
(\ref{sextic3})-(\ref{sextic5}) they reduce to the well known formulas
describing the solution of the simple harmonic oscillator in terms of 
wavefunction zeros \cite{UshRev,Cal2}. In this case the number $M$ can
be considered a free parameter because it is not any longer
correllated with the form of the potential.
 
\section{Evolution equations}
 
Let us next consider the time-dependent Schroedinger equation
\begin{equation}
\label{evolution1}
  i\hbar \frac{\partial \Psi (x,t)}{\partial t}=
  \left\{-\frac{\hbar^2}{2}\frac{\partial^2}{\partial x^2}+
  V(x)\right\}\Psi(x,t).
\end{equation}
for potential (\ref{sextic1}). It is obvious that any linear combination of
the stationary solutions
\begin{eqnarray}
\label{evolution2} 
  \Psi(x,t)=\sum_{n=1}^{N_{sol}} c_{n}\Psi_{n}(x)
  e^{\scriptstyle -iE_{n}t/\hbar}=\nonumber \\
  =\left(\sum_{n=1}^{N_{sol}}c_{n}
  \prod_{i=1}^{M}(x-\xi_{i})
  e^{\scriptstyle -iE_{n}t/\hbar}\right)
  \exp\left[-\frac{1}{\hbar}
  \left(\frac{bx^2}{2}+\frac{ax^4}{4}\right)\right] 
\end{eqnarray}
gives a certain dynamical solution of equation (\ref{evolution1}). 
Remember that the pre-exponential factor in (\ref{sextic3}) is always a 
polynomial of degree $M$. Therefore also the pre-exponential factor in 
(\ref{evolution2}) is a certain polynomial and hence its zeros are functions 
of time. This enables one to write
\begin{equation}
\label{evolution3}
  \Psi (x,t)=C(t)\prod_{i=1}^{M}(x-\xi _i(t))
  \exp \left[-\frac{1}{\hbar} 
  \left(\frac{bx^2}{2}+\frac{ax^4}{4}\right)\right] .
\end{equation}
 
Formula (\ref{evolution3}) enables one to derive the evolution equations
for the functions $\xi _i(t)$ and $C(t)$. 
For this it is convenient to rewrite equation (\ref{evolution1}) in the form
\begin{equation}
\label{dynamics1}
  V(x)=i\hbar \frac \partial {\partial t}\ln \Psi(x,t)+
  \frac{\hbar^2}{2}\left\{\left(\frac \partial {\partial x}
  \ln \Psi(x,t)\right)^2+\frac{\partial^2}{\partial x^2}
  \ln \Psi(x,t)\right\} .
\end{equation}
Substituting expressions (\ref{sextic1}) and (\ref{evolution3}) into equation
(\ref{dynamics1}) we obtain after some algebra the relation
\begin{eqnarray}
  i\hbar \frac{\dot{C}(t)}{C(t)}-
  \hbar a \sum_{i=1}^{M} \xi_{i}^{2}(t)-
  \hbar a x \sum_{i=1}^{M} \xi_{i}(t)-
  \hbar b \left( M+\frac{1}{2} \right)+ \nonumber \\
\label{dynamics2}
  -\sum_{i=1}^{M}\frac{\hbar}{x-\xi_{i}(t)}\left( i \dot{\xi}_{i}(t)-
  \sum_{k=1,k\neq i}^{M} \frac{\hbar}{\xi_{i}(t)-\xi_{k}(t)}+
  b \xi_{i}(t)+ a \xi_{i}^{3}(t)\right)=0 
\end{eqnarray}
from which it immediately follows that the functions $\xi _i(t)$ must 
satisfy the following system of first order differential equations
\begin{equation}
\label{dynamics3}
  i\dot{\xi_i(t)}=\sum_{k=1,k\neq i}^{M} 
  \frac{\hbar}{\xi_i(t)-\xi_k(t)}-b\xi_i(t)-a\xi_i^3(t),\quad i=1,\dots ,M
\end{equation}
supplemented by the condition
\begin{equation}
\label{dynamics4}
  \sum_{i=1}^{M}\xi_i(t)=0.
\end{equation}
For the function $C(t)$ we obtain on the other hand
\begin{equation}
\label{dynamics5}
  C(t)=\exp \left[-i\left(a\sum_{i=1}^{M}\int \xi_i^2(t)dt+
  b\left(M+\frac{1}{2}\right) t \right) \right] .
\end{equation}
The system (\ref{dynamics3})-(\ref{dynamics4}) is the dynamical extension of 
the stationary system (\ref{sextic5})-(\ref{sextic6}). 
Since the functions $\xi_i(t)$ are assumed to be complex, 
we face a typical example of a complex dynamical system of dimension
$M$. It is remarkable that this system can be rewritten in the 
following ``potential'' form
\begin{equation}
\label{dynamics6} 
  i\dot\xi_i(t)=-\frac{\partial}{\partial \xi_i(t)}U(\xi(t)),
\end{equation}
where
\begin{equation}
\label{dynamics7}
  U(\xi )=-\hbar \sum_{k=1,k\neq i}^{M} 
  \ln (\xi_i(t)-\xi_k(t))+\frac{b}{2}\sum_{i=1}^{M}\xi_i^2(t)+
  \frac{a}{4}\sum_{i=1}^{M}\xi_i^4(t)
\end{equation} 
is playing the role of a complex potential.

\section{Complex multi-particle systems}
 
Despite the fact that system (\ref{dynamics6}) cannot be directly
derived from the Lagrange principle, it is possible to relate it to a
certain Lagrangian system in the following way. For this, consider the
complex ``Lagrangian'' 
\begin{equation}
\label{complex1}
  L(\xi,\dot{\xi})=\frac{1}{2}\sum_{i=1}^{M}\dot{\xi_i}^2(t)-
  \frac{1}{2}W(\xi)
\end{equation}
with
\begin{equation}
\label{complex2}
  W(\xi)=\sum_{i=1}^{M}
  \left(\frac{\partial U(\xi)}{\partial \xi_i}\right)^2.
\end{equation}
It is not difficult to see that this Lagrangian (\ref{complex1}) can be
represented in the form
\begin{equation}
\label{complex3}
  L(\xi,\dot{\xi})=-\frac{1}{2}\sum_{i=1}^{M}\Lambda_i^2(\xi,\dot{\xi})+
  i\frac{d}{dt}U(\xi),
\end{equation}
where
\begin{equation}
\label{complex4}
  \Lambda_i (\xi,\dot{\xi})=
  i\dot{\xi_i}(t)-\sum_{k=1,k\neq i}^{M}
  \frac{\hbar}{\xi _i(t)-\xi_k(t)}+
  b\xi_i(t)+a\xi_i^3(t).
\end{equation}
The total time derivative in (\ref{complex3}) can be omitted because it 
does not affect the form of the equations of motion. 
The form of these equations derived from the action principle reads then
\begin{equation}
\label{complex5}
  \sum_{i=1}^M\left[ \frac{d}{dt}\left( \Lambda_i(\xi,\dot{\xi})
  \frac{\partial \Lambda_i(\xi ,\dot{\xi})}{\partial \dot{\xi}_n}\right)
  -\Lambda_i(\xi,\dot{\xi}) 
  \frac{\Lambda_i(\xi,\dot{\xi})}{\partial \xi_n}\right]=0,
  \quad n=1,\dots,M.
\end{equation}
From (\ref{complex5}) it immediately follows that the solutions  
\begin{equation}
\label{complex6}
  \Lambda_i(\xi,\dot{\xi})=0,\quad i=1,\dots,M 
\end{equation}
are automatically solutions of the dynamical equations for the Lagrangian
(\ref{complex1}). Note that system (\ref{complex6}) exactly coincides with 
equations (\ref{dynamics3}). To derive the form of the Lagrangian 
(\ref{complex1}) it is sufficient to substitute expression 
(\ref{dynamics7}) into (\ref{complex1}). This gives 
\begin{eqnarray}
  L(\xi,\dot{\xi})=
  \frac{1}{2} \sum_{i=1}^{M}\dot{\xi}_{i}^{2}(t)-
  \sum_{i=1}^{M} \sum_{k=1,k>i}^{M}
  \frac{\hbar^2}{(\xi_{i}(t)-\xi_{k}(t))^2}+\nonumber \\
\label{complex7}  
  -\sum_{i=1}^{M}\frac{\xi_{i}^2(t)}{2}\left(a\xi_{i}^2(t)+b\right)^2+
  \hbar a\left(M-\frac{3}{2}\right)\sum_{i=1}^{M}\xi_{i}^{2}(t).
\end{eqnarray}
We have obtained a complex Lagrangian system supplemented with the
additional constraint (\ref{dynamics4}). This system can obviously be 
regarded a deformation of a complexified version of the famous 
Calogero-Moser system with Lagrangian
\begin{equation}
\label{complex8}
  L_0(\xi,\dot{\xi})=\frac{1}{2}\sum_{i=1}^{M}
  \dot{\xi_i}^2(t)-\sum_{i=1}^{M} \sum_{k=1,k>i}^{M}
  \frac{\hbar ^2}{(\xi_i(t)-\xi_i(t))^2}-
  \frac{b^2}{2}\sum_{i=1}^{M}\xi_i^2(t).
\end{equation}
Note that the above construction is very similar to the one 
usually used in constructing the soliton solutions of some classical 
dynamical equations. Therefore it is natural to interpret the
solutions of equation (\ref{complex6}) describing the motion of 
wavefunction zeros in quantum QES model (\ref{sextic1}) as solitons 
of the classical multi-particle system with Lagrangian
(\ref{complex1}). 
The relation between the undeformed theories of the simple harmonic 
oscillator and the Calogero-Moser system is well known in the 
literature \cite{Cal1,Cal2}.

\section{Classical matrix model of the sextic anharmonic oscillator}
 
Let $X$ be a $M\times M$ traceless complex matrix whose entries are
considered as dynamical variables of a certain complex dynamical system. 
The Lagrangian of this system can be chosen in the form
\begin{equation}
\label{classical1}
  {\cal L}(X,\dot{X})=\frac{1}{2}\mbox{Tr} \dot{X}^2-\mbox{Tr}V(X)
\end{equation}
with $V(X)$ given by formula (\ref{sextic1}). The corresponding dynamical
equations then read
\begin{equation}
\label{classical2}
  \ddot{X}=-\frac{\partial}{\partial X}V(X).
\end{equation}
It is known that any complex matrix with non-coinciding eigenvalues can 
be reduced to diagonal form by means of an appropriate similarity
transformation
\begin{equation}
\label{classical3}
  X=S\Xi S^{-1}
\end{equation}
where we used the notation 
$\Xi =\mbox{diag}\{\xi_1(t),\dots,\xi_M(t)\}$ with 
$\xi_1(t)+\cdots +\xi_M(t)=0$. Taking the time derivative of 
(\ref{classical3}) we obtain
\begin{equation}
\label{classical4}
  \dot{X}=SLS^{-1},
\end{equation}
where
\begin{equation}
\label{classical5}
  L=\dot{\Xi}+[M,\Xi]
\end{equation}
and
\begin{equation}
\label{classical6}
  M=S^{-1}\dot{S}.
\end{equation}
Differentiating (\ref{classical4}) once more and using (\ref{classical2})
and (\ref{classical3}) we get the equation
\begin{equation}
\label{classical7}
  \dot{L}+[M,L]=-\frac{\partial}{\partial \Xi} V(\Xi)
\end{equation}
which together with (\ref{classical5}) is equivalent to system (\ref
{classical2}). Note that (\ref{classical7}) has the form of a deformed 
Lax equation. A relation between equations (\ref{classical2}) and 
(\ref{classical7}) can also be established in the opposite direction: 
Assume that we have two matrices $M$ and $L$ satisfying
(\ref{classical7}) and (\ref{classical5}). Let $S$ be a solution of 
the linear evolution equation
\begin{equation}
\label{classical8}
  \dot{S}=SM.
\end{equation}
Then the function $X$ in (\ref{classical3}) is a solution of equation 
(\ref{classical2}). 
It is easy to check that the matrices $L$ and $M$ with components
\begin{eqnarray}
  L_{ii}=\dot{\xi}_{i}(t),  \quad
  L_{ik}=\frac{i \hbar}{\xi_{i}(t)-\xi_{k}(t)},
  \nonumber\\
  M_{ii}=\sum_{k=1, k \neq i}^{M}
  \frac{-i \hbar}{(\xi_{i}(t)-\xi_{k}(t))^2}, \quad
  M_{ik}=\frac{-i \hbar}{(\xi_{i}(t)-\xi_{k}(t))^2}
  \label{classical9}
\end{eqnarray}
satisfy the system (\ref{classical5}), (\ref{classical7}) provided the
functions $\xi _i(t)$ are solutions of the deformed Calogero-Moser 
system (\ref{complex7}). 
As mentioned above for this it is sufficient that $\xi _i(t)$
be solutions of system (\ref{dynamics3})-(\ref{dynamics4}) describing 
the motion of the wavefunction zeros in the quantum QES model with 
potential (\ref{sextic1}).\\
From the above reasonings it follows that we essentially established a
relationship between the quantum QES model with Hamiltonian
\begin{equation}
\label{classical10}
  H=\frac{p^2}2+\frac{x^2}2\left( ax^2+b\right) ^2-\hbar
  a\left( M+\frac 32\right) x^2
\end{equation}
and the classical matrix model with Hamiltonian
\begin{equation}
\label{classical11}
  {\cal H}=\mbox{Tr}\left( \frac{P^2}2+\frac{X^2}2\left(
  aX^2+b\right) ^2-\hbar a\left( M-\frac 32\right) X^2\right) .
\end{equation}

\section{The large $M$ limit}
 
It is remarkable that the hamiltonians of both the quantum and classical 
models (\ref{classical10}) and (\ref{classical11}) essentially coincide.
The difference is of order $\hbar a.$ Let us demonstrate now that this 
difference also disappears if we take a proper large $M$ limit of 
these models.\\
The neccessity for taking the limit $M\rightarrow \infty $ in formulas
(\ref{classical10}) and (\ref{classical11}) arises from the following
reasonings. For finite $M$ it is not correct to speak of an equivalence 
of models (\ref{classical10}) and (\ref{classical11}). The point is that
for finite $M$ model (\ref{classical10}) describes the evolution
of only a certain finite superposition of quantum states (because the model
(\ref{classical10}) is quasi-exactly solvable). However, if $M$ tends to 
infinity, then the number of stationary  states in this superposition also
tends to infinity and fills all the spectrum of the model. Only in this case
we can say that models (\ref{classical10}) and (\ref{classical11})
are equivalent. But how to proceed to the large $M$ limit? It is clear 
that if we simply take $M=\infty $ in formulas (\ref{classical10}) and
(\ref{classical11}) then we obtain a meaningless (minus) infinity. In order
to get a finite expression we must take into account that the parameters $a$ 
and $b$ may also depend on $M.$ Choosing this dependence according to 
the conditions
\begin{equation}
  ab=g,\quad \frac{b^2}{2}-\hbar a M=\frac{\omega^2}{2}
\end{equation}
in which $g$ and $\omega $ are fixed (positive) numbers, we find that in
the large $M$ limit the parameter $a$ must behave as $a\sim M^{-1/3}.$ This
means that in the limit $M\rightarrow \infty $ the terms containing
$a^2$ and $a$ (i.e. the sextic term and also the harmonic term
responsible for the difference between the models) disappear and we 
obtain the two models
\begin{equation}
  H=\frac{p^2}{2}+\frac{\omega^2 x^2}{2}+gx^4
\label{limit1}
\end{equation}
and 
\begin{equation}
  {\cal H}=\mbox{Tr}\left(\frac{P^2}{2}+\frac{\omega^2 X^2}{2}+
  gX^4\right)
\label{limit2}
\end{equation}
with exactly coinciding classical and quantum hamiltonians.
Note however that model (\ref{limit1}) is a
one-particle quantum model while model (\ref{limit2}) is a classical
model of traceless infinite matrices. The models are equivalent in the
sense that the soliton-like solutions in the classical model 
(\ref{limit2}) describe the evolution of the wavefunctions in the
quantum model (\ref{limit1}) and vice verso.
 
\section{Discussion}
 
When speaking of a relationship between quantum and classical mechanics one
usually thinks of a pair of quantum and classical models related to each
other by a certain quantization -- dequantization procedure. The
hamiltonians of these models, considered as functions of coordinates and
momenta, formally coincide up to terms of order $\hbar $. However, the
mathematical meaning of the coordinates and momenta is essentially different
in the quantum and classical case. In the quantum case they are operators in
Hilbert space (infinite matrices) while in the classical case -- they are 
simply numbers. Correspondingly, the number of quantities 
(degrees of freedom) neccessary to fix uniquely the state of a
dynamical system is infinitely larger in the quantum case compared to 
the classical one. In this sense the  transition to the classical 
limit is equivalent to freezing infinitely many  degrees of freedom 
of the quantum system.\\
There are however several examples where the correspondence between the
quantum and classical model is not approximate but exact and is not  related
to any limiting procedure and any lose of degrees of freedom. 
The idea underlying such examples is based on a proper parametrization
of the wavefunction by an infinite number of parameters depending on
time and considered as canonically conjugated dynamical variables. 
If such a parametrization is found then we can associate with a given 
quantum model a certain classical one which in this case
should neccessarily be infinite-dimensional. 
It is quite clear that a priori there are no  reasons for any relation
between the forms of the corresponding quantum and classical
hamiltonians. In general, they may be of absolutely different
nature. Consider a simple but most instructive example. Let us take 
the evolution equation for a certain quantum model with hamiltonian $H$
\begin{equation}
  i\hbar \partial_t\Psi=H\Psi
\label{con1}
\end{equation}
Since the wavefunction is complex, $\Psi =Q+iP$ , we can rewrite 
equation (\ref{con1}) in real form
\begin{equation}
  \hbar \partial_tQ=HP,\quad \hbar \partial_tP=-HQ,
\end{equation}
which, after introducing the functional
\begin{equation}
  {\cal H}(Q,P)=\frac{1}{\hbar} \left(Q,HQ\right)+
  \frac{1}{\hbar} \left(P,HP\right) ,
\end{equation}
can in turn be rewritten in the form of the classical Hamilton-equations:
\begin{equation}
  \partial_tQ=\frac{\delta {\cal H}(Q,P)}{\delta P}\quad 
  \partial_tP=-\frac{\delta {\cal H}(Q,P)}{\delta Q}.
\end{equation}
We see that irrespective of the form of the initial quantum model, the
resulting classical one describes an infinite-dimensional coupled 
harmonic oscillator.
$\\ \\$
In this paper we have found examples for an exact relationship between
quantum and classical models. The main distinguished feature of these
examples is that the potentials of the initial quantum and the resulting
classical models exactly coincide. In this case the construction of the
classical counterpart of a given quantum model (in our case the quartic or
harmonic oscillator) is extremely simple. One should simply replace the
operators of coordinate and momentum entering into the quantum model by
infinite traceless complex matrices and, after this, take the trace of this
matrix hamiltonian. What we obtain will be just the hamiltonian of
a classical model whose solutions contain the complete information of the
dynamics of the wavefunctions in the initial quantum model. 
The origin of this coincidence is not clear to us at the moment but it
is quite obvious that it cannot be accidental. 
It would be tempting to conjecture that any one-dimensional quantum 
model with hamiltonian
\begin{equation}
  H=\frac{p^2}{2}+V(x)
\end{equation}
is somehow equivalent to the classical infinite and traceless complex
matrix model with the same hamiltonian
\begin{equation}
  {\cal H}=\mbox{Tr}\left(\frac{P^2}{2}+V(X)\right)
\end{equation}
We have an idea how to check this conjecture at least for models with
polynomial anharmonicity and hope to publish the results in the near future.


\begin{thebibliography}{99}
\bibitem{UshRev}  A. Ushveridze, {\em Quasi-exactly solvable models in
                  quantum mechanics}, Soviet Journal of Particles and 
                  Nuclei 20 (5), 1989, 504-528 
 
\bibitem{UshLeb}  A. Ushveridze, Sov. Phys. -- Lebedev
                  Inst. Rep. {\bf2}, 50; 54 (1988) 
 
\bibitem{May}  A. Chorin, J. Marsden, {\em A mathematical introduction to 
               fluid mechanics}, Springer Verlag, New York 1993 

\bibitem{Cic1}  G. M. Cicuta, S. Stramaglia, A. Ushveridze, {\em Quartic
                anharmonic oscillator and random matrix theory}, 
                Modern Physics Letters A {\bf 11}, 119-129 (1996) 

\bibitem{Cic2}  G. M. Cicuta, A. Ushveridze, {\em Quasi-exactly
                solvable problems and random matrix theory}, 
                Physics Letters A {\bf 215} 167 (1996) 

\bibitem{Shi}  M. A. Shifman, Int. J. Mod. Phys. A {\bf 4}, 2897 (1989) 


\bibitem{FinKam}  F. Finkel, N. Kamran, {\em Quasi-exactly solvable
                  time-dependent potentials}, 
                  hep-physics/9705022 17 May 1997 

\bibitem{KrajUshWal}  A. Krajewska, A. Ushveridze, Z. Walczak, 
                      {\em Anti-isospectral transformations in quantum
                      mechanics}, Modern Physics Letters A, Vol. 12, 
                      No. 17 (1997), 1225-1234
 
\bibitem{Gaud}  M. Gaudin, {\em La Fonction d'Onde de Bethe}, 
                Paris:  Masson, (1983) 
 
\bibitem{Ush}  A. Ushveridze, {\em Quasi-Exactly Solvable Problems 
               in Quantum Mechanics}, IOP Publishing, 1994 
 
\bibitem{Cal1}  F. Calogero, {\em Solvable many-body problems and 
                related mathematical findings}, in C. Bardos and 
                D. Bessis (eds.), {\em Bifurcation Phenomena in Mathematical
                Physics and Related Topics}, D. Reidel Publishing
                Company 1980, 371-384 

\bibitem{Cal2}  F. Calogero, {\em Motion of poles and zeros of special
                solutions of nonlinear and linear partial differential
                equations and related ``solvable'' many-body problems},
                Il Nuovo Cimento, Vol. 43 B, No. 2, 177-241 (1978)

\bibitem{Olsh1} M. A. Olshanetsky, A. M. Perelomov, {\em Explicit
                  solution of the Calogero model in the classical case
                  and geodesic flows on symmetric spaces of zero
                  curvature}, Lett. Nuovo Cim. {\bf 16} 333-339 (1976) 
\end{thebibliography}
\end{document}